\documentclass[english,12pt]{article}
\usepackage[T1]{fontenc}
\usepackage[latin1]{inputenc}
\usepackage{amsmath}
\usepackage{amsthm}
\usepackage{amssymb}
\usepackage{graphicx}

\makeatletter
 \theoremstyle{plain}    
 \newtheorem{thm}{Theorem}

 \numberwithin{equation}{section} 

 \numberwithin{figure}{section} 

 \theoremstyle{plain}

 \theoremstyle{remark}

 \theoremstyle{plain}    
 
 \theoremstyle{plain}

 \theoremstyle{plain}

 \theoremstyle{plain}

 \theoremstyle{plain}    

 \theoremstyle{plain}    
 \newtheorem{lem}{Lemma} 
 
 \theoremstyle{definition}

 \theoremstyle{definition}
  \newtheorem{example}{Example}
        
\usepackage{babel}
\makeatother

\def\={\overset{def}{=}}
\def\R{\mathrm{I\kern-.2emR\kern-.1em}}
\def\N{\mathrm{I\kern-.2emN\kern-.03em}}

\begin{document}

\newcommand{\te}{\rightarrow}
\newcommand{\RE}{\bf R}
\newcommand{\CE}{\bf C}
\newcommand{\ZE}{\bf Z}
\newcommand{\al}{\alpha}
\newcommand{\be}{\beta}
\newcommand{\de}{\delta}
\newcommand{\ep}{\epsilon}
\newcommand{\Om}{\Omega}
\newcommand{\om}{\omega}
\newcommand{\ga}{\gamma}
\newcommand{\Ga}{\Gamma}
\newcommand{\subs}{\subset}
\newcommand{\pa}{\partial}

\newcommand{\lab}{\left | }
\newcommand{\rab}{\right | }
\newcommand{\lnm}{\left\|}
\newcommand{\rnm}{\right\|}
\newcommand{\lob}{\left (}
\newcommand{\rob}{\right )}
\newcommand{\lcb}{\left [}
\newcommand{\rcb}{\right ]}
\newcommand{\lcub}{\left \{ }
\newcommand{\rcub}{\right \} }

\newcommand{\D}{\displaystyle}

\addtolength{\jot}{2.5mm}

\newlength{\lnbrk}
\setlength{\lnbrk}{2.5mm}

\title{Geometric Properties and Non-blowup of 3-D Incompressible Euler Flow
\thanks{This work was
supported in part by NSF under the grant DMS-0073916 and ITR Grant
ACI-0204932.}}
\author{Jian Deng\thanks{Applied and Comput. Math, Caltech, Pasadena, CA 91125.
Email: jdeng@acm.caltech.edu}
\and Thomas Y. Hou\thanks{Applied and Comput. Math, Caltech, Pasadena, CA 91125. Email: hou@acm.caltech.edu}
\and Xinwei Yu\thanks{Applied and Comput. Math., Caltech, Pasadena, CA 91125. 
Email: xinweiyu@acm.caltech.edu}} 

\maketitle

\abstract{By exploring a local geometric property of the vorticity field along
a vortex filament, we establish a sharp relationship between the geometric 
properties of the vorticity field and the maximum vortex stretching. This new 
understanding leads to an improved result of the global existence of the 3-D 
Euler equation under mild assumptions that are consistent with the observations 
from recent numerical computations.}

\section{Introduction}

The problem of global existence/blowup of smooth solutions for the
three-dimensional incompressible Euler flow, which is governed by
the 3-D Euler equations:

\begin{eqnarray}\label{3deuler}
\left. \begin{array}{rcl}
u_t+(u\cdot\nabla)u & = & \nabla p\\
\nabla\cdot u & = & 0\\
u\mid_{t=0} & = & u_{0}
\end{array} \right.
\end{eqnarray}
is a long time outstanding question. It plays a very important role
in understanding the core problems in hydrodynamics such as the onset
of turbulence. Much effort has been made in both theory and numerics
trying to answer this question, see, e.g. Beale-Kato-Majda \cite{B.K.M}, 
Caflisch \cite{Caf}, Constantin-Fefferman-Majda \cite{C.F.M}, and 
Babin-Mahalov-Nicolaenko \cite{B.M.N}. Through these efforts, it is 
realized that the above issue is closely related to the stretching of 
the vorticity $\omega\equiv\nabla\times u$, which is governed by the 
following evolution equation:

\begin{eqnarray}
\begin{array}{rcl}
        \D\frac{D\omega}{Dt} & = & (\nabla u)\omega \label{euler}\\[\lnbrk]
        \omega\mid_{t=0} & = & \omega_{0}=\nabla\times u_{0},
\end{array}
\end{eqnarray}
where 
        $\frac{D}{Dt}\equiv\partial_{t}+(u\cdot\nabla)$ 
is the material derivative. In a well-known paper by Beale, Kato
and Majda (\cite{B.K.M}), it has been shown that the smooth solution 
$u(x,t)$ for the 3-D Euler flow blows up at $t=T$ if and only if 
$\int_{0}^{t}\|\omega(\cdot,s)\|_{\infty}ds\nearrow\infty$
as $t\nearrow T$. Some variants and improvements have appeared
in the last two decades. Very recently, Ogawa and Taniuchi \cite{O.T.} 
have shown that the above $L^\infty$ norm estimate on vorticity in 
Beale-Kato-Majda's blow-up criterion can be replaced by a weaker BMO 
norm estimate. 

The above result implies that we should study the dynamic growth of vorticity
in the flow. It has been observed from the early 80s in the last century
that small vortex tubes dominate the vorticity field in later times of the 
flow, especially in near-singular situations. This observation gives 
impetus on studying the details of the evolution of regions in which 
vorticity concentrates. People have been trying to find conditions on 
the geometry of the vorticity field that can be used to
exclude blow-up by rigorous
mathematical proofs. In particular, Constantin, Fefferman and Majda 
\cite{C.F.M} prove that if there is up to time $T$ an $O(1)$ region 
in which the vorticity vector 
$\xi(x,t)\equiv\frac{\omega(x,t)}{\left|\omega(x,t)\right|}$
is smoothly directed, i.e., the maximum norm of $\nabla \xi$
in this region is $L^2$ integrable in time from $0$ to $T$, and 
the maximum norm of velocity in some $O(1)$ neighborhood of this
region is uniformly bounded in time, 
then no blow-up can occur in this region up to time $T$. 
Another way of attacking the problem is by taking advantage of 
incompressibility and is explored by 
Cordoba and Fefferman in \cite{C.F}, in which the possibility of 
uniform collapsing of vortex tubes with $O(1)$ length 
that don't twist or bend violently are ruled out under the assumption 
that the infinity norm of velocity in a neighborhood of the region
under consideration is integrable in time. 

Many numerical computations have been performed to search for
possible candidates for a finite time blow-up. Some of the better
known examples are given by Kerr \cite{K,K.1,K.2,K.3}, Pelz 
\cite{P.R, P.R.1}, and
Grauer-Marliani-Germaschewski \cite{G.M.G}. Up to now, the most 
probable candidate is the anti-parallel vortex tube setting,
which has been carefully studied by Kerr and others taking advantage
of the ever-growing computing power (see e.g. \cite{K, K.1,K.2, K.3}). 
The magnitudes of maximum vorticity observed in all of these numerical 
computations can be fitted by a growth rate of $(T-t)^{-1}$ 
which is the critical case in the Beale-Kato-Majda blow-up criterion. 
Although many numerical results suggesting finite time 
blow-ups have been obtained, 
no conclusive claim has been drawn so far. One thing that is worth 
mentioning is that, in all these computations, it is observed that 
vorticity is concentrated in small regions that are shrinking with time, 
and the shrinkage rate is related to some inverse power of the maximum 
vorticity.

There is still little overlap of the cases studied by the theoretical
and numerical groups, although many results have been obtained and efforts 
made. All the existing theorems deal with $O(1)$ regions in which the 
vorticity vector is assumed to have some regularity, while in numerical 
computations, the regions which have such regularity and contain maximum
vorticity are all shrinking with time. In this article, we try to narrow 
this gap by 
considering cases that are consistent with the numerical observations. 
We prove that no finite time blow-up can occur if some mild assumptions
on the geometric properties of the vorticity vector and the behavior
of the velocity field are satisfied. 

The key to our analysis is the following understanding. The magnitudes 
of vorticity at any two points on one vortex line are related to each
other by the geometry of the vorticity field through the incompressibility
condition. This understanding has not been seen in 
the existing literature 
and is a key to our analysis. Another key factor to our work is the 
reformulation of the problem into a vortex filament setting. Unlike 
previous vorticity growth formulas, this formulation reveals the anisotropic 
nature of vortex stretching and enables us to obtain an improved
global existence result for the incompressible 3-D Euler equation.

Specifically, we obtain two results. The first one says that if the
divergence of the vorticity vector, $\nabla \cdot \xi$, 
along the vortex line segment $[s_1,s_2]$ containing the point of 
maximum vorticity is integrable, i.e., 
\begin{equation}
\lab\int_{s_1}^{s_2} (\nabla \cdot \xi ) ds \rab 
  \le C(T), \quad 0 \le t \le T, 
\label{reg-ori}
\end{equation}
where $s$ is the arc length variable, and $s_2 > s_1$, then no point 
singularity is possible up to time $T$. If the vorticity blows up at
one point within this vortex line segment, then the vorticity must 
blow up simultaneously {\it at the same rate} in the {\it entire} 
vortex line segment. 

The usefulness of the first result lies in the fact that the 
weakly regular orientedness condition expressed by (\ref{reg-ori}) 
is extremely localized. It is a condition along one local vortex 
line segment. Since people have observed numerically some form 
of partial regularity of the vorticity vector in a small inner region 
containing the point of maximum vorticity, we can readily apply this 
criterion to check the validity of some singularity scenarios reported
in numerical computations. For example, in Pelz's computation 
\cite{P.R, P.R.1}, a tube-shaped region of length scale $(T-t)^{1/2}$ 
is highlighted as a candidate for a finite time blow-up. A simple
calculation shows that the criterion (\ref{reg-ori}) is satisfied within 
this inner tube-shaped region. This casts doubt on the validity of Pelz's 
claim on the finite time formation of a point singularity. To validate 
Pelz's claim, one needs to perform a more careful numerical study to 
check whether there exists a nonvanishing vortex line segment within 
which condition (\ref{reg-ori}) is satisfied or whether the vorticity 
within the inner tube-shaped region blows up at the same rate.

Our second theorem proves the global existence of the incompressible 
3-D Euler equation under some relatively mild assumptions. In this
theorem, we deal with the case when the length of the weakly regularly
oriented vortex line segment can shrink to zero as the time approaches
to the alleged singularity time. It gives a sharper dynamic description 
of the vortex stretching. Assume that at each time $t$
there exists some vortex line segment $L_t$ on which the 
maximum vorticity is comparable to the global maximum vorticity. 
We denote by $L(t)$ the arc length of this vortex line segment.
In addition to satisfying a variant of (\ref{reg-ori}), we assume 
that $L(t) \|\kappa \|_{L^\infty(L_t)}$ (here $\kappa$ is curvature of 
the vortex line $L_t$) is bounded, and that the maximum norms of the 
normal and tangential velocity components along the local vortex 
line segment $L_t$ are integrable in time. The length of the local 
vortex line segment, $L(t)$, is allowed to shrink to zero as time 
approaches to the alleged singularity time. Under these 
assumptions, we can prove that no finite time blow-up is possible. 
To simplify our analysis and to obtain a concrete rate of shrinkage of 
$L(t)$, we present a slightly weaker version of the result in this 
paper by assuming an upper bound of the growth rate in time of the 
normal and tangential velocity components along $L_t$. 

Our second theorem to some extent improves the previous results obtained 
by Constantin-Fefferman-Majda \cite{C.F.M} and Cordoba-Fefferman \cite{C.F}. 
First of all, our result requires a very localized and weaker
assumption on the regularity of the vorticity vector $\xi$. In \cite{C.F.M}, 
the gradient of the vorticity vector is assumed to be $L^2$ integrable in 
time in an $O(1)$ region containing the maximum vorticity.
In contrast, we only assume that the divergence of the vorticity vector 
is integrable along a local vortex line segment and  
$L(t) \|\kappa \|_{L^\infty(L_t)}$ is bounded. 
The length of the vortex line segment, 
$L(t)$, can shrink to zero as the time approaches to the alleged singularity
time. The numerical computations by Kerr \cite{K} and Pelz \cite{P.R}
have demonstrated that there is indeed a small region in which 
vorticity attains its global maximum and the vorticity vector has some
partial regularity.  However, the size of this region shrinks 
rapidly to zero in a rate proportional to some inverse power of maximum 
vorticity. Thus there is a significant gap between the assumption
on the smoothly directed region in \cite{C.F.M} and what has been
observed numerically. On the other hand, our assumption on the
partial regularity of the vorticity vector is very mild and localized
along one vortex line segment so that we can apply our result to 
check the validity of some numerical studies, such as those by
Kerr (\cite{K,K.1,K.2,K.3}), in which finite time singularity of the
3-D Euler equation has been alleged.

It is also worth mentioning that in our second theorem, we only assume 
that the normal and tangential velocity components within the local 
vortex line segment $L_t$ are integrable in time. In comparison, the 
maximum norm of the entire velocity field within an $O(1)$ region is 
assumed to be bounded in \cite{C.F.M}. In the case of the collapse of 
a regular vortex tube, the maximum velocity is given by the rotational
component of the velocity field in the cross section normal to the 
direction of the vortex tube. As the vortex tube collapses, the 
rotational component of the velocity field may blow up proportional 
to the square root of the maximum vorticity from Kelvin's circulation
theorem. The normal velocity component generally corresponds to the
speed of the motion of the vortex tube, which may remain bounded
even in the collapse of the vortex tube. On the other hand, we 
expect that the tangential velocity
component is smaller than the maximum velocity field 
due to the cancellation of vorticity vectors in the inner region, 
leading to one order reduction of the velocity kernel.

We would like to emphasize that our analysis reveals a close connection 
between the global existence of the 3-D Euler equation and the local 
geometric property of a vortex line segment containing the maximum 
vorticity. This observation sheds useful light in our future effort 
in studying the dynamical interplay between the local geometric property 
of the vortex filament and the maximum vortex stretching.

This paper is organized as follows. We highlight our main results in 
Section 2 and describe their implications by applying
them to study some recent numerical computations. In Section 3,
we explore the geometry of the vorticity field and the incompressibility 
condition in depth and prove our two main theorems. 

\vspace{0.2in}
\noindent
{\bf Notations}

Throughout this paper, we will reserve some characters for some particular quantities
according to the following rules of notations: 

\begin{itemize}
        \item $C$ or $c$: generic constants, whose value may change from line to 
line. When not otherwise indicated, the values of $C(c)$ are independent of any 
of the data. 
        \item $\xi$ is always the direction of vorticity vectors, i.e., 
$\xi\equiv \omega/\lab\omega\rab$. 
        \item $T$ will always denote the alleged time when a finite time blow-up occurs. 
\end{itemize}

We will also use the following notations for convenience: 

\begin{itemize}
        \item $\sim$: We write $a(t)\sim b(t)$ if there are absolute constants 
$c,C>0$ such that $ c\lab a(t)\rab \le \lab b(t)\rab \le C\lab a(t) \rab $. 
        \item $\gtrsim$($\lesssim$): We write $a(t)\gtrsim b(t)$ if there is 
an absolute constant $c>0$ such that $\lab a(t)\rab \ge c \lab b(t) \rab $. 
$a(t)\lesssim b(t)$ is defined similarly.
\end{itemize}

\section{Main Results and Their Implications}

In this section, we present our two main results. The first one says that
if the divergence of the vorticity vector along a nonvanishing local vortex line 
segment containing the maximum vorticity is integrable in time, then no point 
singularity is possible. If the vorticity blows up at one point, then the vorticity 
along this vortex line segment must blow up simultaneously at the singularity time. 
Our second result gives a sharp criterion for the dynamic blow-up of vorticity. 
With additional assumptions on the curvature of the local vortex line and the
growth rate of the normal and tangential velocity components along the vortex
line, we prove that no blowup is possible in finite time. Below we describe
these two results and discuss how they can be applied to check validity of 
some numerical studies in which singularities of the 3-D Euler equation have 
been alleged.

Our first theorem is as follows:

\begin{thm}\label{theorem1}
        We consider any 3-D incompressible flow ( Euler or Navier-Stokes ). Let $x(t)$ be
        a family of points such that
        $\lab\omega(x(t),t)\rab \gtrsim \Omega(t)\equiv \lnm\omega(\cdot,t)\rnm_{L^\infty(\R^3)}$. Assume that for
        all $t\in [0,T)$ there is another point $y(t)$ on the same vortex line as $x(t)$,
        such that the direction of vorticity
        $\xi(x,t)\equiv \frac{\omega(x,t)}{\lab\omega(x,t)\rab} $
        along the vortex line between $x(t)$ and $y(t)$ is well-defined. If
we further assume that
        
\begin{equation}
\lab \int_{x(t)}^{y(t)} (\nabla\cdot\xi)(s,t)\ ds \rab \le C 
\label{reg-ori1}
\end{equation}
 for some absolute constant $C$, and
        $$ \int_0^T \lab\omega(y(t),t)\rab\ dt < \infty, $$
        then there will be no blow-up up to time $T$. Moreover, we have
\begin{equation}
e^{-C}\le \frac{\lab\omega(x,t)\rab}{\lab\omega(y,t)\rab}\le e^{C} .
\end{equation}

\end{thm}

The proof of Theorem 1 is quite simple and will be deferred to Section 3.

The above theorem gives a practical criterion on judging possible blow-up in a numerical
computation. It also suggests that, when searching for a 
finite time blow-up numerically, one has to pay attention to the 
geometric property of vortex filaments. It
is not enough to just track the maximum vorticity magnitude and the point at which this
maximum is attained. The vorticity magnitudes at other points are also crucial. In
particular, the above theorem implies that if there is a nonvanishing vortex line segment 
containing the maximum vorticity up to time $T$ such that (\ref{reg-ori1}) is satisfied, 
then no point singularity is possible up to this time $T$. To illustrate, we apply Theorem 
\ref{theorem1} to the numerical results of Pelz \cite{P.R, P.R.1}.

\begin{example}\label{pelz}
In \cite{P.R, P.R.1}, Pelz studied a class of incompressible flows with strong 
symmetry and conjectured that such flows can lead to a finite time blow-up. 
In these computations, vorticity is 
concentrated in small vortex tubes of length scale $\sim (T-t)^{1/2}$. After a 
re-scaling $x\mapsto (T-t)^{-1/2}x$, these tubes seem to have a regular shape. 
This suggests that the length of this inner region scales like
$(T-t)^{1/2}$ and the scaling of $\nabla \cdot \xi $ within this
inner region is of the order $(T-t)^{-1/2}$. Let us take the point 
$x(t)$ to be the point inside one tube where the maximum vorticity is attained, 
and $y(t)$ to be a point on the same vortex line, but outside the tube.
It is easy to check that within this inner region, condition (\ref{reg-ori1})
is satisfied. By Theorem \ref{theorem1} we see that if the maximum vorticity 
{\it outside} these small tubes is integrable in time, then there is no 
blow-up inside the tubes. It is likely that the maximum vorticity
outside these small tubes has a growth rate smaller than that inside these
small regions. This casts doubt on the validity of Pelz's claim on the 
finite time formation of a point singularity. To validate Pelz's claim, 
one needs to perform more careful numerical study to check whether 
there exists a nonvanishing vortex line segment within which condition 
(\ref{reg-ori}) is satisfied or whether the vorticity within the inner
tube-shaped region blows up at the same rate.  
\end{example}

Our second result is concerned with the dynamic blow-up of one vortex line.
As in \cite{B.K.M}, we assume that the initial velocity field, $u_0$, is
smooth and vanishes rapidly at infinity, more specifically 
$u_0 \in H_0^{7/2}(\R^3)$.
Denote by $\Omega(t)$ the maximum vorticity in the whole 3-D space. We consider
a family of vortex line segments $L_t$ along which maximum vorticity is comparable 
to $\Omega(t)$. Denote by $L(t)$ the arc length of $L_t$, 
$U_{\xi}(t)\equiv \max_{x,y\in L_t} \lab (u\cdot\xi)(x,t)-(u\cdot\xi)(y,t) 
\rab$, 
$ U_n(t)\equiv \max_{L_t} \lab u\cdot n\rab $, and
$ M(t)\equiv \max (\| \nabla\cdot\xi \|_{L^\infty(L_t)}, \| \kappa \|_{L^\infty(L_t)})$
where $\kappa$ is the curvature of the vortex line and $n$ is the unit normal
vector of $L_t$. Further, we denote by $X(\alpha ,t)$ the Lagrangian flow map
\cite{C.M}.

Now we can state our second theorem.

\begin{thm}\label{theorem2}
Assume there is a family of vortex line segments $L_t$ and $T_0\in [0,T)$, 
such that $X(L_{t_1},t_2) \supseteq L_{t_2}$ for all $T_0<t_1<t_2<T$. We also
assume that $\Omega(t)$ is monotonely increasing and 
$\|\omega (t) \|_{L^\infty(L_t)} \ge c_0\Omega(t)$ for some $c_0 >0$ when $t$ is 
sufficiently close to $T$. Furthermore, we assume that
\begin{enumerate}
        \item $\lcb U_{\xi}(t)+ U_n(t)M(t)L(t)\rcb \lesssim (T-t)^{-\alpha}$ for some $\alpha\in (0,1)$,
        \item $M(t)L(t) \le C_0$, and
        \item $L(t)\gtrsim (T-t)^\beta$ for some $\beta<1-\alpha$,
\end{enumerate}
then there will be no blow-up in the 3D incompressible Euler flow up to time $T$.
\end{thm}

The proof of Theorem 2 relies on the geometric property of the 3-D Euler
equation in a crucial way and will be deferred to Section 3. Here we would like
to make a few remarks on the assumptions of the theorem and discuss how one
can use this result to check the validity of some alleged 3-D Euler singularities 
obtained by numerical computations.

First, we remark that the first two assumptions of Theorem 2 
are quite natural. From numerical computations, it has been observed that
incompressible flows at later times are dominated by small regions of large 
magnitude of vorticity that shrink in all three directions in the Eulerian 
coordinates. In fact, they should shrink in the Lagrangian coordinates
as well. To see this, we argue that
if they don't shrink in the Lagrangian coordinates, the volumes of these 
small regions 
would be non-decreasing since any Lagrangian region carried by the 
flow must maintain its volume due to the incompressibility of the flow. 
Thus these small regions must have at least one stretching direction along
which the small regions grow in the Eulerian coordinates. This contradicts 
with the observation that these small regions shrink in all three 
directions. Now that these small regions shrink in all directions in the
Lagrangian coordinates, it is reasonable to assume that there is one 
Lagrangian point $X(\alpha,t)$ that is contained in all these regions. 
Now if we take $L_t$ to be the vortex line segment that passes $X(\alpha,t)$, 
then these two assumptions would be satisfied. Note that the assumption 
$M(t)L(t) \le C$ is a sufficient condition to satisfy (\ref{reg-ori1}). 

Next, we note that in Theorem \ref{theorem2}, we used $U_\xi(t)+U_n(t)M(t)L(t)$
instead of the more observable quantity $U(t)\equiv \max_{L_t}\lab u(\cdot,t)\rab $.
This is because we believe that $U_\xi(t)+U_n(t)M(t)L(t)$ may grow slower than
$U(t)$. To see this, we first consider the term $U_\xi(t)$, which is defined as 
the maximum of the difference between the tangential velocity at any two points
on $L_t$. In the case of collapsing vortex tubes, it is likely that the
the tangential velocity has a better regularity along vortex lines than along
the direction normal to vortex lines. In this case, the term $U_\xi (t)$
can be much smaller than the velocity itself. 
Even if such regularity is not available, we can bound this term by 
$2\max_{L_t}\lab u\cdot\xi \rab$. 
By the Biot-Savart law, we have
$$
        u\cdot\xi(x,t)=\frac{1}{4\pi}\int_{\R^3} \frac{y}{\lab y\rab^3} \times
                        \omega(x+y,t) \cdot \xi(x,t)\ dy.
$$
If vorticity is concentrated in a small region around $x$, and 
$\xi$ has some regularity within this small region,
then there will be an extra order of cancellation at $y=0$ in the
integral kernel for the tangential velocity.
Therefore we should expect $u\cdot\xi$ to be smaller than $|u|$.

We now consider the term $U_n(t)M(t)L(t)$. In a regular vortex tube, if the
maximum vorticity is achieved at the center vortex line, then $U_n$ should
correspond to the velocity of the motion of the vortex tube in the direction
normal to itself. Even in the case of the vortex tube collapsing, the speed of
the motion of the vortex tube itself is usually bounded. In this case,
the maximum velocity component should come from the rotational component
of the velocity. A formal argument based on Kelvin's circulation theorem
shows that the rotational velocity component is of the order $\Omega(t)^{1/2}$.
More generally, we can estimate the maximum velocity in terms of the maximum 
vorticity as follows: 
\begin{equation}
U(t)\lesssim \Omega(t)^{3/5}, 
\label{velvor}
\end{equation}
where $U(t)$ and $\Omega(t)$ are the maximum velocity and maximum vorticity
in the whole space respectively. This estimate will be proved rigorously
in Lemma \ref{3_5bound} in the Appendix without making any regularity 
assumption on the vorticity vector. Therefore, as long as 
$\Omega(t) \lesssim (T-t)^{-5/3+\epsilon}$ for arbitrary small 
$\epsilon >0$, then we will have 
$U(t) \lesssim  (T-t)^{-1+\epsilon}$. 
In fact,  in almost all numerical 
computations so far, $\Omega(t)\lesssim (T-t)^{-1}$. Thus, we can reasonably 
expect that $U(t)\lesssim (T-t)^{-3/5}$. 

Theorem \ref{theorem2} can be viewed as a refinement of some existing theoretical 
results, and it is also more applicable to numerical observations. We illustrate 
these points through the following examples.

\begin{example}\label{CFM}
First we review the theorem by Constantin-Fefferman-Majda \cite{C.F.M}. In
\cite{C.F.M}, Constantin, Fefferman, and Majda prove that no finite time blow-up can 
occur under two main assumptions: (i) 
$\int_0^T \lnm \nabla\xi \rnm_{L^\infty(W_t)}^2 dt < \infty$, where 
$W_t \equiv X(W_0,t)$ with $W_0$ being an $O(1)$ Lagrangian region at $t=0$; and
(ii)
$\lnm u(\cdot,t) \rnm_{L^\infty(W_t)}$ is bounded by some absolute constant $U$
(For technical reasons, they also make some additional assumptions).
Now if we further assume that $\lnm \nabla\xi\rnm_{L^\infty(W_t)}$ has
a growth rate at $t\rightarrow T$, then their two main assumptions turn into
$$ M(t) \ll (T-t)^{-1/2}, \quad \quad U_\xi(t)+U_n(t) \lesssim 2U. $$
Since $W_t$ is carried by the flow, we can take $L_t$ to be any vortex line 
in $W_t$. In particular,
we can take $L_t$ such that $L(t)\sim (T-t)^{1/2}$. This is equivalent
to taking $\alpha=0, \beta=1/2$
in Theorem 2. We see that the three conditions in Theorem \ref{theorem2} are all 
satisfied and there will be no finite time blow-up.  

Theorem 2 to some extent improves the previous result obtained
by Constantin-Fefferman-Majda \cite{C.F.M}. First of all, our result requires a 
weaker and very local assumption on the regularity of the vorticity vector $\xi$ 
along one vortex line segment. In \cite{C.F.M}, the maximum norm of the gradient of 
the vorticity vector is assumed to be $L^2$ integrable in time in an $O(1)$ region
containing the maximum vorticity, and the maximum norm of the velocity field is 
required to be bounded in this $O(1)$ region. In contrast, we essentially
assume that the divergence of the vorticity vector and the curvature are
integrable along a local vortex line segment whose length can shrink 
to zero as the time approaches to the alleged singularity
time. The fact that the size of this local weakly regularly oriented region can 
shrink to zero with appropriate rate enables us to essentially eliminate the gap 
between our theoretical result and what has been observed numerically, which is 
significant. 
\end{example}

\begin{example}\label{CF}
Next we review the main result by Cordoba-Fefferman \cite{C.F}. There they consider a fixed cube region
$Q\equiv I_1\times I_2 \times I_3$, and a vortex tube $\Omega_t$ that only intersects with $\partial Q$ 
twice at any time $t$, one in the upper face, and one in the lower face. Furthermore, $\lnm u(\cdot,t)\rnm_{L^\infty}$
is assumed to be integrable from $0$ to $T$. Under these assumptions,
they prove that the volume of $\Omega_t$ can
not shrink to $0$ as $t\nearrow T$.

If we assume that $M(t)$ is bounded by some constant, and $U(t)$ has a algebraic growth rate as $t\nearrow T$,
then we can get the result in \cite{C.F} by Theorem \ref{theorem2} by taking $L(t)\sim 1$.
Note that we can take $L(t)\sim 1$ since the cube region $I_1\times I_2 \times I_3$ is fixed.
In fact, in this case Theorem \ref{theorem2} rules out general blow-ups, while the result in
\cite{C.F} only rules out the possibility that the whole vortex tube shrink to one filament.
\end{example} 

\begin{example}\label{Kerr}
Finally we apply our result to the numerical computations by Kerr. In a 
sequence of papers \cite{K,K.1, K.2,K.3}, Kerr 
observed that when $t$ is close enough to the alleged blow-up time $T$, the region bounded by 
the contour of $0.6\times$maximum vorticity has the length scale
$(T-t)^{1/2}$ in the vortex line direction, and is contained in a box with
length scale $(T-t)^{1/2}$ in all three directions. Within this region,
vortex lines are ``relatively straight'' (\cite{K.2}), except that in a smaller 
inner region they have curvature $\sim (T-t)^{-1/2}$ (\cite{K.1}). It is
also observed that the maximum velocity inside this region remains bounded up to 
time $T$(\cite{K.2}). Thus we can take $L(t)\sim (T-t)^{1/2}$, 
$M(t)\lesssim (T-t)^{-1/2}$, and $U_\xi(t),U_n(t)\sim (T-t)^0$. The assumptions
in Theorem \ref{theorem2} are satisfied. Therefore it is quite possible that there 
will be no blow-up in this case. We plan to perform a more careful numerical
study to further investigate this possible blow-up scenario by verifying the
above scaling of various geometric and flow properties.
\end{example}

\section{Proofs of Theorem 1 and Theorem 2}

In this section, we prove our two main theorems in this paper. We first 
prove Theorem 1. Before we present the proof of Theorem 1, we first study
the properties of the vorticity field. The key to our analysis is the
incompressibility condition. It turns out that, when combined with the 
geometrical properties of the vorticity field, this condition becomes a 
constraint on the behavior of the flow, thus an obstacle for a finite time 
blow-up to occur.

\subsection{Direction and magnitude of vorticity}

It has long been observed that at later times incompressible flows are dominated
by small vortex tubes in which the vorticity concentrates. This phenomenon has also 
been observed in recent numerical computations ( Kerr \cite{K,K.1,K.2,K.3},
Pelz \cite{P.R, P.R.1} ). A vortex tube is a collection of vortex lines, so it is
natural to study the behavior of the magnitudes of vorticity along one vortex line. 

First, we have the following lemma, which relates, through the incompressibility 
condition, the vortex line geometry to the magnitude of vorticity. 

\begin{lem}\label{first_lem}
Let $\xi(x,t) \stackrel{def}= \frac{\omega(x,t)}{\lab \omega(x,t) \rab}$ be the direction of 
the vorticity vector. Assume at a fixed time $t>0$ the vorticity $\omega(x,t)$ is $C^1$ in $x$.
We denote 
$$N = \{ x \in R^3: \omega(x,t) \neq 0 \}.$$ 
Then at this time $t$, for any $x \in N$ there holds 
\begin{equation}\label{dyn}
\frac{\partial \lab \omega \rab}{\partial s}(x,t) = - (\nabla \cdot \xi(x,t)) \lab \omega \rab (x,t). 
\end{equation}
where $s$ is the arc length variable along the vortex line passing $x$. 
Further more for any $y$ that is on the same vortex line segment as $x$, (\ref{dyn}) 
then gives
\begin{equation}\label{arcest}
\lab \omega(y,t) \rab = \lab \omega(x,t) \rab \cdot e^{\int_x^y (-\nabla \cdot \xi)(s,t) ds},
\end{equation}
as long as the vortex line segment connecting $x$ and $y$ lies in $N$, where the 
integration is along the vortex line. 

\end{lem} 

\begin{proof}
Notice that $\omega = \lab \omega \rab \xi$. Since $\omega(x,t) \neq 0$, $\xi\equiv \frac{\omega}{\lab \omega \rab}$ is  well defined in a neighborhood of $x$. The incompressibility condition $\nabla \cdot \omega = 0$ then gives

\begin{eqnarray}
\left.
\begin{array}{rcl}
0=\nabla \cdot \omega & = & \nabla \cdot (\lab \omega \rab \xi) \\
        &=& (\nabla \lab \omega \rab) \cdot \xi + (\nabla \cdot \xi) \lab \omega \rab \\
        &=& (\xi \cdot \nabla) \lab \omega \rab + (\nabla \cdot \xi) \lab \omega \rab.
\end{array}
\right.
\label{eqn:incom}
\end{eqnarray}
It is easy to check that the directional derivative $\xi \cdot \nabla$
is actually the arc length derivative along the vortex line, i.e.
$\xi \cdot \nabla = \frac{\partial}{\partial s}$. 
Therefore we obtain from (\ref{eqn:incom}) that

\begin{equation}\label{dyn2}
\frac{\partial \lab \omega \rab}{\partial s} = - (\nabla \cdot \xi) \lab \omega \rab,
\end{equation}
with $s$ being the arc length variable. 
Equation (\ref{arcest}) then follows from integrating (\ref{dyn2}) along the vortex line. 
\end{proof}

Now we are ready to present a simple proof of Theorem 1.

\vspace{0.1in}
\noindent
{\bf Proof of Theorem 1:}
Using (\ref{arcest}) and the assumption that
$ \int_0^T \lab\omega(y(t),t)\rab\ dt < \infty $ we obtain 
$$ \int_0^T \lab\omega(x(t),t)\rab\ dt <\infty. $$
Then by our assumption on $\omega (x(t),t)$, we have 
$$ \int_0^T \Omega(t)\ dt \lesssim 
   \int_0^T \lab\omega(x(t),t)\rab\ dt < \infty .
$$
Thus the theorem follows from the Beale-Kato-Majda theorem \cite{B.K.M}. 
The estimate on the ratio of $\omega(x,t)$ and $\omega(y,t)$ is a direct 
consequence of (\ref{arcest}). This completes the proof of Theorem 1. 

\subsection{Stretching of Vortex Lines}

Before we prove Theorem 2, we need to study how the relative rate of arc length 
stretching along a vortex filament is related to the relative rate of maximum
vorticity growth in time.

For any starting time $t_1$ and some time $t>t_1$, consider the evolution of a vortex
line. Let $s$ and $\beta$ be the arc length parameter of this vortex line at time $t$ and
$t_1$ respectively. We can write, for this very vortex line, $s=s(\beta,t)$. Note
that $s(\beta,t_1)=\beta$. Then we have the following lemma. 

\begin{lem}
For any point $\alpha$ at time $t_1$ such that $\omega(\alpha,t_1)\neq 0$, 
let $X(\alpha,t)$ be the position of the same particle at time
$t\ge t_1$. Then we have

\begin{equation}
\frac{\partial s}{\partial\beta}(X(\alpha,t),t)=
        \frac{\lab \omega(X(\alpha,t),t)\rab}{\lab \omega(\alpha,t_1) \rab } . 
\end{equation}
\end{lem}
\begin{proof}
Note that in our notation, $X(\alpha,t_1) =\alpha$.
It is well known that for $3$-D Euler flows we have \cite{C.M}
$$
\omega(X(\alpha,t),t)=\nabla_\alpha X(\alpha,t) \cdot \omega(\alpha,t_1) . 
$$
Then we obtain
\begin{eqnarray*}
        \lab\omega(X(\alpha,t),t)\rab 
                &=& \xi(X(\alpha,t),t)\cdot\omega(X(\alpha,t),t) \\
                &=& \xi(X(\alpha,t),t)\cdot \nabla_\alpha X(\alpha,t)\cdot \xi(\alpha,t_1) \lab\omega(\alpha,t_1)\rab . 
\end{eqnarray*}
Note that $\xi(X(\alpha,t),t)=\frac{\partial X(\alpha,t)}{\partial s} $ for any $t$, where $s$
is the arc length variable of the vortex line that passes $X(\alpha,t)$ at time $t$. We can
further simplify the above equations as

\begin{eqnarray*}
        \lab\omega(X(\alpha,t),t)\rab 
                &=& \D\frac{\partial X(\alpha,t)}{\partial s} \cdot \nabla_\alpha X(\alpha,t)
                                \cdot \D\frac{\partial\alpha}{\partial\beta} \lab \omega(\alpha,t_1)\rab \\
                &=& \D\frac{\partial X(\alpha,t)}{\partial s}\cdot
                                \D\frac{\partial X(\alpha,t)}{\partial \beta} \lab \omega(\alpha,t_1)\rab \\
                &=& (\D\frac{\partial X(\alpha,t)}{\partial s}\cdot 
                                \D\frac{\partial X(\alpha,t)}{\partial s})
                                \D\frac{\partial s}{\partial \beta}\lab \omega(\alpha,t_1)\rab \\
                &=& \lab \xi(X(\alpha,t),t)\rab^2 \D\frac{\partial s}{\partial \beta}\lab \omega(\alpha,t_1)\rab \\
                &=& \D\frac{\partial s}{\partial \beta}\lab \omega(\alpha,t_1)\rab . 
\end{eqnarray*}
This completes the proof of Lemma 2.
\end{proof}

It is well-known that the evolution of the magnitude of vorticity along any particle path is governed by
the following equation (\cite{C}),

\begin{equation}
 \frac{D\lab \omega(x,t)\rab}{Dt} = \xi(x,t)\cdot(\nabla u(x,t)\cdot\xi(x,t))\lab \omega(x,t) \rab ,
\end{equation}
where 
$\frac{D}{Dt}\equiv \partial_t + u\cdot\nabla $ is the material derivative. 
Then the above lemma immediately gives the equation that governs the 
arc length stretching $s_\beta$.
If we denote $x=X(\alpha,t)$, then we have

\begin{eqnarray}\label{stretching}
\left.
\begin{array}{rcl}
        \D\frac{D s_\beta}{Dt}(x,t))&=& \lcb \xi\cdot(\nabla u\cdot\xi) \rcb s_\beta \\
                &=& \lcb (\xi\cdot\nabla u)\cdot \xi \rcb s_\beta \\
                &=& \lcb (\xi\cdot\nabla)(u\cdot\xi) - u\cdot (\xi\cdot\nabla)\xi \rcb s_\beta \\
                &=& \lcb (u\cdot\xi)_s - \kappa (u\cdot n)\rcb s_\beta \\
                &=& (u\cdot\xi)_\beta - \kappa (u\cdot n)s_\beta ,
\end{array}
\right.
\end{eqnarray}
where we have used the fact $\xi\cdot\nabla = \frac{\partial}{\partial s}$ and the 
well-known basic relation in differential geometry

\begin{equation}\label{def_of_n}
        \frac{\partial \xi}{\partial s}=\kappa n , 
\end{equation}
with $\kappa=\lab \xi\cdot\nabla \xi\rab $ being the curvature and $n$ the unit 
normal vector of the vortex line. 

Now we integrate equation (\ref{stretching}) along the vortex line:

\begin{eqnarray}\label{stretching_1}
\left.
\begin{array}{rcl}
        \D\frac{D[s(\beta_2,t)-s(\beta_1,t)]}{Dt}&=&(u\cdot\xi)(X(\beta_2,t),t)-
(u\cdot\xi)(X(\beta_1,t),t) \\
                &&{}-\D\int_{\beta_1}^{\beta_2} \kappa(X(\eta,t),t)\cdot(u\cdot n)
       s_\eta\ d\eta .
\end{array}
\right.
\end{eqnarray}
Further, we integrate (\ref{stretching_1}) over some time interval $[t_1,t]$. We get

\begin{eqnarray}\label{stretching_2}
\left.
\begin{array}{rcl}
        s(\beta_2,t)-s(\beta_1,t)&=&s(\beta_2,t_1)-s(\beta_1,t_1) \\
                &&{}+\D\int_{t_1}^t \lob (u\cdot\xi)(X(\beta_2,\tau),\tau)-(u\cdot\xi)
                (X(\beta_1,\tau),\tau)\rob \ d\tau \\
                &&{}-\D\int_{t_1}^t \int_{\beta_1}^{\beta_2} \kappa(\eta,\tau)\cdot 
                (u\cdot n)s_\eta\ d\eta d\tau .
\end{array}
\right.
\end{eqnarray}
Let $l(t)\equiv s(\beta_2,t)-s(\beta_1,t)>0$ and denote by $l_{12}$ the 
vortex line segment connecting the points $X(\beta_1,t)$ and $X(\beta_2,t)$. 
It follows from (\ref{stretching_2}) that 

\begin{eqnarray}\label{stretching_3}
\left.
\begin{array}{rcl}
l(t)&\le &l(t_1)+\D\int_{t_1}^t 
 (\lab (u\cdot\xi)(X(\beta_2,\tau),\tau) -
(u\cdot\xi)(X(\beta_1,\tau),\tau)\rab ) d\tau \\
                &&      {}+ \D\int_{t_1}^t M(\tau)\| u\cdot n\|_{L^\infty (l_{12})}
(\tau) l(\tau)\ d\tau, 
\end{array}
\right.
\end{eqnarray}
where 
$M(t) = \max(\|\nabla \cdot \xi \|_{L^\infty (l_{12})},\|\kappa\|_{L^\infty (l_{12})})$. 
Inequality (\ref{stretching_3}) reveals how the stretching of vortex lines is controlled by
the velocity field and the geometry of the vorticity field. Furthermore, we will derive
an inequality to bound the relative ratio of the magnitudes of vorticity at different 
time by the relative ratio of the arc lengths of the vortex lines. This provides
a sharp estimate on the growth rate of vorticity in terms of the arc length stretching 
of vortex lines.

\begin{lem}\label{lem_str_omega}
Let $l_t$ be a vortex line segment that is carried by the flow. Denote its length by $l(t)$, and let 
$M(t)$ be defined as in Theorem 2. Then for any point $X(\alpha',t)\in l_t$, 
we have
\begin{equation}\label{stretching_omega}
e^{-(M(t)l(t)+M(t_1)l(t_1))}\frac{\lab \omega(X(\alpha',t),t) \rab}{\lab \omega(\alpha',t_1)\rab}
\le \frac{l(t)}{l(t_1)}
\le e^{(M(t)l(t)+M(t_1)l(t_1))}\frac{\lab \omega(X(\alpha',t),t) \rab}{\lab \omega(\alpha',t_1)\rab}.
\end{equation}
\end{lem}

\begin{proof}
Let $\beta$ denote the arc length parameter at time $t_1$. Denote by $l_t$ 
the vortex line segment from $0$ to $\beta$, and use $s$ as the arc length parameter 
at time $t$. Now by the mean value theorem and Lemma 2 we have

\begin{eqnarray*}
        \frac{l(t)}{l(t_1)}=\frac{\D\int_0^\beta s_\beta(\eta)\ d\eta}{\beta}= 
                 s_\beta(\eta') = \frac{\lab \omega(X(\alpha'',t),t) \rab}{\lab \omega(\alpha'',t_1)\rab},
\end{eqnarray*}
for some $\alpha''$ on the same vortex line. Now (\ref{stretching_omega})
follows from Lemma 1. This completes the proof of Lemma 3.
\end{proof}

By combining (\ref{stretching_omega}) and (\ref{stretching_3}), we obtain
\begin{equation}\label{stretching_4}
        \begin{array}{rcl}
                \lab \omega(X(\alpha,t),t)\rab &\le& 
                                                e^{(M(t)l(t)+M(t_1)l(t_1))}\lab\omega(X(\alpha,t_1),t_1)\rab \\
                                                && \cdot \lcb 1 + \D\frac{C}{l(t_1)}
                                                 \D\int_{t_1}^t 
\left ( U_{\xi}(\tau) + M(\tau)U_{n}(\tau) l(\tau)\right ) \ d\tau \rcb,
        \end{array}
\end{equation}
for any $X(\alpha,t)$ that lies in $l_t$. Let
$\Omega_l(t) = \|\omega(\cdot ,t)\|_{L^\infty(l_t)}$. We can easily derive from 
(\ref{stretching_4}) the following inequality: 

\begin{equation}\label{stretching_5}
                        \Omega_l(t) \le 
                                                e^{(M(t)l(t)+M(t_1)l(t_1))}\Omega_l(t_1) \lcb 1 + \D\frac{C}{l(t_1)}
                                                 \D\int_{t_1}^t 
\left ( U_{\xi}(\tau) + M(\tau)U_{n}(\tau) l(\tau) \right ) d\tau \rcb.
\end{equation}
Inequality (\ref{stretching_5}) shows how the growth of vorticity is controlled by the properties
of the flow. This inequality is important to our analysis in our proof of 
Theorem \ref{theorem2} and will be used heavily. 

\subsection{Interplay between the Geometry and Growth Rate}

This subsection is devoted to the proof of Theorem \ref{theorem2}. 

\medskip

\noindent
{\bf Proof of Theorem 2:}
We prove Theorem 2 by contradiction.
First, by translating the initial time we can assume that the assumptions
in Theorem 2 hold in $[0,T)$. 
Define 
\begin{equation}
r\equiv (R/c_0)+1,
\label{eqn:r}
\end{equation} 
where $R\equiv e^{2C_0}$, $C_0$ is the 
constant in the theorem such that $M(t)L(t)\le C_0$ for all 
$t\in [0,T)$, and $c_0$ is the constant such that 
$\Omega_L(t)\ge c_0 \Omega(t)$. Throughout the proof we denote
$\Omega_L(t) \equiv \|\omega(\cdot ,t)\|_{L^\infty(L_t)}$.
The reason for choosing the parameter $r$ this way will become
clear later in the proof. 
If there were a finite time blow-up at time $T$, we would have
$$ \int_0^T \Omega(t)\ dt = \infty, $$
or equivalently for any $t_1\in [0,T)$, 
$$ \int_{t_1}^T \Omega(t)\ dt =\infty. $$
Then necessarily we have 
$\Omega(t)\nearrow \infty$ as $t\nearrow T$. Now we can take
a time sequence $t_1,t_2,\ldots,t_n,\ldots$ such that 
\begin{equation}
\Omega(t_{k+1})=r\Omega(t_k), 
\label{eqn:Omek}
\end{equation}
where $r$ is defined in (\ref{eqn:r}).
Since $\Omega(t)$ is monotone, and $T$ is the 
smallest time such that $\int_0^T \Omega(t)\ dt=\infty$, 
it is obvious that $t_n\nearrow T$ as $n\rightarrow \infty$. 

Now we choose $l_{t_2} = L_{t_2}$. By our assumptions on $L_t$,
there is $l_{t_1}\subset L_{t_1}$ such that $X(l_{t_1},t_2)=l_{t_2}$. 
If we further denote 
$$ 
        \Omega_l(t_i) \equiv \lnm \omega(\cdot,t)\rnm_{L^\infty(l_{t_i})}
        \quad i=1,2,
$$
then by taking $t=t_2$ in (\ref{stretching_omega}) we would have
\begin{eqnarray*}
  l(t_1)& \ge & 
    l(t_2)\frac{1}{R}
    \frac{\lab\omega(\alpha',t_1) \rab}{\lab\omega(X(\alpha',t_2),t_2)\rab}\\
    &\ge& l(t_2)\frac{1}{R^2}\frac{\Omega_L(t_1)}{\Omega_L(t_2)} ,
\end{eqnarray*}
where the last inequality is due to the assumption $M(t)L(t)\le C_0$ and 
Theorem \ref{theorem1}. Note that by assumption we have
$\Omega_L(t)\ge c_0 \Omega(t)$. Thus $l(t_1)$ can be further bounded 
from below by
\begin{eqnarray}\label{lt1Lt2}
  \begin{array}{rcl}
  l(t_1) &\ge& l(t_2)\D\frac{c_0}{R^2}\frac{\Omega(t_1)}{\Omega(t_2)} \\ \\
         &=& Cl(t_2)=CL(t_2)\gtrsim (T-t_2)^\beta ,
  \end{array}
\end{eqnarray}
where $C=\frac{c_0}{R^2 r}$ is independent of time. 

On the other hand, we have from (\ref{stretching_5})
\begin{equation}\label{Omegalt2}
                        \Omega_l(t_2) \le 
                                                e^{(M(t_2)l(t_2)+M(t_1)l(t_1))}\Omega_l(t_1) \lcb 1 + \D\frac{C}{l(t_1)}
                                                 \D\int_{t_1}^{t_2} 
( U_{\xi}(\tau) + M(\tau) U_{n}(\tau) l(\tau) ) \ d\tau \rcb.
\end{equation}
By the assumption of Theorem \ref{theorem2}, we have 
\begin{eqnarray*}
  M(t_2)l(t_2)+M(t_1)l(t_1) &\le & C_0 \\
  U_\xi(\tau)+U_n(\tau)M(\tau)l(\tau) 
     &\lesssim & (T-\tau)^{-\alpha}.
\end{eqnarray*}
Then it follows from (\ref{Omegalt2}) and (\ref{lt1Lt2}) that
$$
        \Omega_l(t_2) \le R\Omega_l(t_1)+C\D\frac{\Omega_l(t_1)}{(T-t_2)^\beta}
                \D\int_{t_1}^{t_2} (T-\tau)^{-\alpha}\ d\tau .
$$
Note that the constant $C$ here depends on $R$, $r$ and $c_0$. 

Applying our assumption that $\Omega_L(t)\ge c_0\Omega(t)$, we have
\begin{eqnarray*}
  \Omega(t_2) &\le & \frac{1}{c_0}\Omega_L(t_2) 
          = \frac{1}{c_0}\Omega_l(t_2) \\
          &\le & \frac{R}{c_0}\Omega_l(t_1)+\frac{C}{c_0}
            \D\frac{\Omega_l(t_1)}{(T-t_2)^\beta}
               \D\int_{t_1}^{t_2} (T-\tau)^{-\alpha}\ d\tau \\
           &\le & \frac{R}{c_0}\Omega(t_1)+\frac{C}{c_0}
            \D\frac{\Omega(t_1)}{(T-t_2)^\beta}
               \D\int_{t_1}^{t_2} (T-\tau)^{-\alpha}\ d\tau \\
           &\le & (r-1) \Omega(t_1)+\frac{C}{(1-\alpha)c_0}
               \frac{\Omega(t_1)}{(T-t_2)^\beta}
               \lcb (T-t_1)^{1-\alpha} - (T-t_2)^{1-\alpha} \rcb,
\end{eqnarray*}
where $r=(R/c_0)+1$ is defined in (\ref{eqn:r}). 
We still denote $C/(c_0(1-\alpha ))$ by $C$. The generic constant $C$ 
now depends on $R$, $r$, $c_0$ and is proportional to $(1-\alpha)^{-1}$. 
Since $(T-t_2)^{1-\alpha}>0$, we can discard it and obtain 
\begin{equation}\label{Omegat_2_t_1}
    \Omega(t_2)\le (r-1)\Omega(t_1)+C\Omega(t_1)
               \frac{(T-t_1)^{1-\alpha}}{(T-t_2)^\beta}.
\end{equation}
Since $\Omega(t_2)=r\Omega(t_1)$, we can cancel $\Omega(t_1)$ from
both sides of (\ref{Omegat_2_t_1}) and obtain
$$
  r\le (r-1)+C\frac{(T-t_1)^{1-\alpha}}{(T-t_2)^\beta},
$$
which gives
$$ 
        (T-t_2)^\beta \le C(T-t_1)^{1-\alpha},
$$
or equivalently
$$ 
        (T-t_2)\le C(T-t_1)^{1+2\delta},
$$
where
$$ 
        2\delta \equiv \frac{1-\alpha}{\beta}-1.
$$
Now it is quite clear that why we take 
$\Omega(t_2)/\Omega(t_1) = r >R/c_0$ and choose 
$r =  (R/c_0)+1$.

Now note that $t_1$ is independent of $C$ and $\delta$, so we can take $t_1$ close enough to $T$ such that
$C(T-t_1)^{\delta} < 1$. Then we have
$$ 
        (T-t_2) \le (T-t_1)^{1+\delta}.
$$

By doing the same thing to each pair $(t_k,t_{k+1})$, we get
$$
        (T-t_{k+1})\le (T-t_k)^{1+\delta}\le (T-t_1)^{(1+\delta)^k}.
$$
If we take $(T-t_1)<1$, this reduces to 
\begin{equation}\label{k_to_kp1}
        (T-t_{k+1})\le (T-t_1)^{1+k\delta} = (T-t_1)\lob(T-t_1)^\delta\rob^k.
\end{equation}

Now we study the condition of $\int_{t_1}^T \Omega(t)\ dt=\infty$
more carefully. By the assumption
that $\Omega(t)$ is monotonely increasing, we have
\begin{eqnarray*}
\int_{t_1}^T \Omega(t)\ dt &=& \sum_{k=1}^\infty \int_{t_k}^{t_{k+1}} \Omega(t)\ dt \\
                                &\le & \sum_{k=1}^\infty \Omega(t_{k+1})(t_{k+1}-t_k) \\
                                &=& \Omega(t_1)\sum_{k=1}^\infty r^k(t_{k+1}-t_k),
\end{eqnarray*}
where we have used $\Omega(t_{k+1})=r\Omega(t_k)=r^k\Omega(t_1)$. 
Since $\int_{t_1}^T \Omega(t)\ dt = \infty$, we obtain
$$ \sum_{k=1}^\infty r^k (t_{k+1}-t_k) = \infty . $$
From this, we conclude that
\begin{eqnarray*}
   \sum_{k=1}^\infty \sum_{l=0}^{k-1} (r^{l+1}-r^l)(t_{k+1}-t_k) 
  &=& \sum_{k=1}^\infty (r^k-1)(t_{k+1}-t_k) \\
  &=& \sum_{k=1}^\infty r^k(t_{k+1}-t_k) - (T-t_1) \\
  &=& \infty. 
\end{eqnarray*}
Since $r=(R/c_0)+1>1$, all the terms $(r^{l+1}-r^l)(t_{k+1}-t_k)$ 
in the summation are positive. 
We can exchange the order of the summation to get
\begin{eqnarray*}
        \sum_{k=1}^\infty \sum_{l=0}^{k-1} (r^{l+1}-r^l)(t_{k+1}-t_k)
                                &=& \sum_{l=0}^\infty \sum_{k=l+1}^\infty (r^{l+1}-r^l)(t_{k+1}-t_k) \\
                                &=& \sum_{l=0}^\infty (r^{l+1}-r^l)(T-t_{l+1})\\
                                &=& (r-1)\sum_{l=0}^\infty r^l(T-t_{l+1}),
\end{eqnarray*}
which implies
$$ \sum_{k=0}^\infty r^k(T-t_{k+1}) = \infty .$$
By substituting (\ref{k_to_kp1}) into the above equation, we get
\begin{equation}\label{contra}
        \sum_{k=0}^\infty \lcb r(T-t_1)^\delta \rcb^k = \infty.
\end{equation}

Note that we can take $t_1$ arbitrarily close to $T$. In particular,
we can take $t_1$ such that $r(T-t_1)^\delta <1$. This implies 
$ \sum_{k=0}^\infty \lcb r(T-t_1)^\delta \rcb^k < \infty$. This 
contradicts with (\ref{contra}). Therefore we conclude that 
$\int_0^T \Omega(t)dt <\infty$. It then follows from the Beale-Kato-Majda 
blow-up criterion that there will be no blow-up up to time $T$. This completes the
proof of Theorem 2. 

\newpage
\noindent
{\bf Appendix. Estimate of Maximum Velocity by Maximum Vorticity}

\vspace{0.2in}

In this appendix, we prove the following lemma: 

\begin{lem}\label{3_5bound}
Let $u(x,t)$ be the solution to 3-D Euler equations (\ref{3deuler}), and 
$\omega(x,t)\equiv \nabla\times u(x,t)$ be the vorticity. Denote
$\Omega(t)\equiv \lnm \omega(\cdot,t)\rnm_{L^\infty(\R^3)}$ and 
$U(t)\equiv \lnm u(\cdot,t)\rnm_{L^\infty(\R^3)}$. Then the following
inequality holds: 
\begin{equation*}
  U(t)\lesssim \Omega(t)^{3/5}. 
\end{equation*}

\end{lem}
\begin{proof}
By the well-known Biot-Savart law \cite{C.M}, we have

\begin{eqnarray*}
u(x,t)=\frac{1}{4 \pi}\int_{\R^3} \frac{y}{\lab y\rab^3}\times \omega(x+y,t)\ dy.
\end{eqnarray*}
Take a common smooth cut-off function $\chi: \lcub 0 \rcub \cup \R^+ \mapsto [0,1]$ such that
$\chi(r)=1$ for $r\le 1$ and $\chi(r)=0$ for $r\ge 2$. Let $\rho>0$ be a small
positive parameter to be determined later. Then we have

\begin{eqnarray*}
\lefteqn{\lab u(x,t) \rab = \lab \frac{1}{4 \pi}\int_{\R^3} \frac{y}{\lab y\rab^3}\times \omega(x+y,t)\ dy \rab} \\
        &= & \lab \frac{1}{4\pi}\int_{\R^3} \chi(\frac{\lab y\rab}{\rho}) \frac{y}{\lab y\rab^3}\times \omega(x+y,t)\ dy 
                                        +\frac{1}{4 \pi} \int_{\R^3} (1-\chi(\frac{\lab y\rab}{\rho})) \frac{y}{\lab y\rab^3}\times (\nabla\times u(x+y,t))\ dy \rab.
\end{eqnarray*}
Invoking integration by part in the second integral, we have
\begin{eqnarray*}
       \lab u(x,t)\rab  
       &\le & \frac{1}{4\pi}\Omega(t)\int_{\lab y\rab\le 2\rho} \frac{1}{\lab y\rab^2}\ dy \\
        && {}+ C \int_{\lab y\rab\ge \rho} \frac{1}{\lab y\rab^3}\lab u(x+y,t)\rab\ dy \\
        && {} + C' \frac{1}{\rho}\int_{\lab y\rab\ge \rho} \frac{1}{\lab y\rab^2}\lab u(x+y,t)\rab\ dy.
\end{eqnarray*}
Using the polar coordinate in the first integral, and the Schwarz inequality 
in the other two, we obtain
\begin{eqnarray*}
       \lab u(x,t)\rab  
       &\le & C\lcb \Omega(t)\rho + \lob\int_{\lab y\rab\ge \rho} \frac{1}{\lab y\rab^6}\ dy\rob^{1/2}
         + \frac{1}{\rho}\lob \int_{\lab y\rab \ge \rho} \frac{1}{\lab y\rab^4}\ dy\rob^{1/2} \rcb, 
\end{eqnarray*}
where we have used the fact that $\lnm u\rnm_{L^2(\R^3)}$ is conserved
in time \cite{C.M}, i.e. 
$\lnm u \rnm_{L^2(\R^3)} = \lnm u_0 \rnm_{L^2(\R^3)} \le C$. 

Finally we use the polar coordinates in the last two integrals, and get
\begin{eqnarray*}
        \lab u(x,t)\rab 
&\le & C\lcb \Omega(t)\rho + \lob\int_\rho^\infty \frac{1}{r^4}\ dr\rob^{1/2} 
                                        + \frac{1}{\rho}\lob\int_\rho^\infty \frac{1}{r^2}\ dr\rob^{1/2} \rcb \\
        &\le & C\lcb \Omega(t)\rho + \rho^{-3/2} \rcb .
\end{eqnarray*} 
By taking $\rho=\Omega(t)^{-2/5}$, we prove the desired estimate. This
completes the proof of Lemma 4.

\end{proof}

\end{document}